\documentclass[a4paper,11pt]{article}
\usepackage{jinstpub} 
\usepackage{subfigure}

\proceeding{\normalsize{11$^{\text{th}}$ International Workshop on Semiconductor Pixel Detectors for Particles and Imaging}\\
 18$^{\text{th}}$ to 22$^{\text{nd}}$ November 2024\\
 Strasbourg, France}

\title{\boldmath Characterization of CMOS sensor using X-ray irradiation}

 \author[a,1]{A. Vijay,\note{Corresponding author.}}
 \author[a]{P. Behera,}
 \author[a]{T. Chembakan,}
 \author[a]{G. Dash} 
 \affiliation[a]{Indian Institute of Technology Madras, Chennai 600036, India}

\emailAdd{anusree.vijay@cern.ch}

\abstract{Recent advancements in particle physics demand pixel detectors that can withstand increased luminosity in the future collider experiments. In response, MALTA ~\cite{Pern}, a novel monolithic active pixel detector, has been developed with a cutting-edge readout architecture. This new class of monolithic pixel detectors is found to have exceptional radiation tolerance, superior hit rates, higher resolution and precise timing resolution, making them ideally suited for experiments at the LHC. To optimize the performance of these sensors before their deployment in actual detectors, comprehensive electrical characterization has been conducted. This study also includes comparative DAC analyses among sensors of varying thicknesses, providing crucial insights for performance enhancement. For the further understanding of the effect of radiation, the sensors are being exposed to different fluence using high intensity X-ray source.}


\arxivnumber{} 

\begin{document}
\maketitle
\flushbottom

\section{Introduction}
\label{sec:intro}
Precise tracking and vertexing are crucial for any Large Hadron Collider experiment. The Tracker subdetector utilizes hybrid pixel detectors and silicon strip sensors \cite{atlas}, with silicon pixel detectors known for their exceptional spatial resolution, fast response, and radiation hardness. An emerging alternative to the hybrid pixel detector is the Depleted Monolithic Active Pixel Sensor (DMAPS), built with CMOS technology. DMAPS simplifies design by integrating sensing and readout on a single substrate, reducing the material budget while offering faster signal processing and enhanced radiation hardness. The paper summarizes the results of the threshold study done on a prototype of MALTA sensor, which is a DMAPS, implemented in the TowerJazz 180 nm CMOS process.

\section{MALTA}
\label{sec:malta}
MALTA is a highly promising candidate for future collider experiments, offering several distinct advantages. Key features include a small pixel pitch of 36.4 $\mu m^{2}$ \cite{carlos}, which ensures excellent spatial resolution, and a compact collection electrode design that minimizes capacitance. The asynchronous architecture of MALTA eliminates the need for clock distribution across the pixel matrix, enabling efficient low-power operation while mitigating the risk of cross-talk. 
The sensor exhibits a power consumption of 10 mW/$cm^{2}$ for its digital components and 70 mW/$cm^{2}$ for its analog components. It features a pixel matrix comprising 512 × 512 pixels. \\
The MALTA sensor incorporates several performance-enhancing process modifications \cite{Munker}. The Standard Process Modification (STD) introduces a uniform $n^{-}$ layer across the p-type substrate for a consistent depletion layer. The NGAP process incorporates gaps in the $n^{-}$ layer structure, and the XDPW technique includes a deep p-well implant at pixel corners, both designed to enhance charge collection when signal charge is produced at the pixel corners. \\
This study focuses on the characterization of the second-generation MALTA sensor, known as the \textbf{MALTA2} sensor \cite{milu}. The MALTA2 sensor is half the size of its predecessor, featuring a matrix of 224 × 512 pixels, and exhibits reduced random telegraph signal noise. While retaining the asynchronous readout architecture of the MALTA sensor, MALTA2 incorporates significant improvements in slow control and front-end design. \\
The sensor has been fabricated on both high-resistivity epitaxial (EPI) and Czochralski (Cz) substrates. Additionally, MALTA2 employs two distinct doping concentrations for the $n^{-}$ layer, referred to as H-dop and VH-dop. These variations in doping concentration are critical for achieving high radiation hardness. 

\section{Experimental Setup}
\label{sec:expsetup}
\begin{figure}[htbp]
\centering
\subfigure[]{
    \includegraphics[width=.4\textwidth]{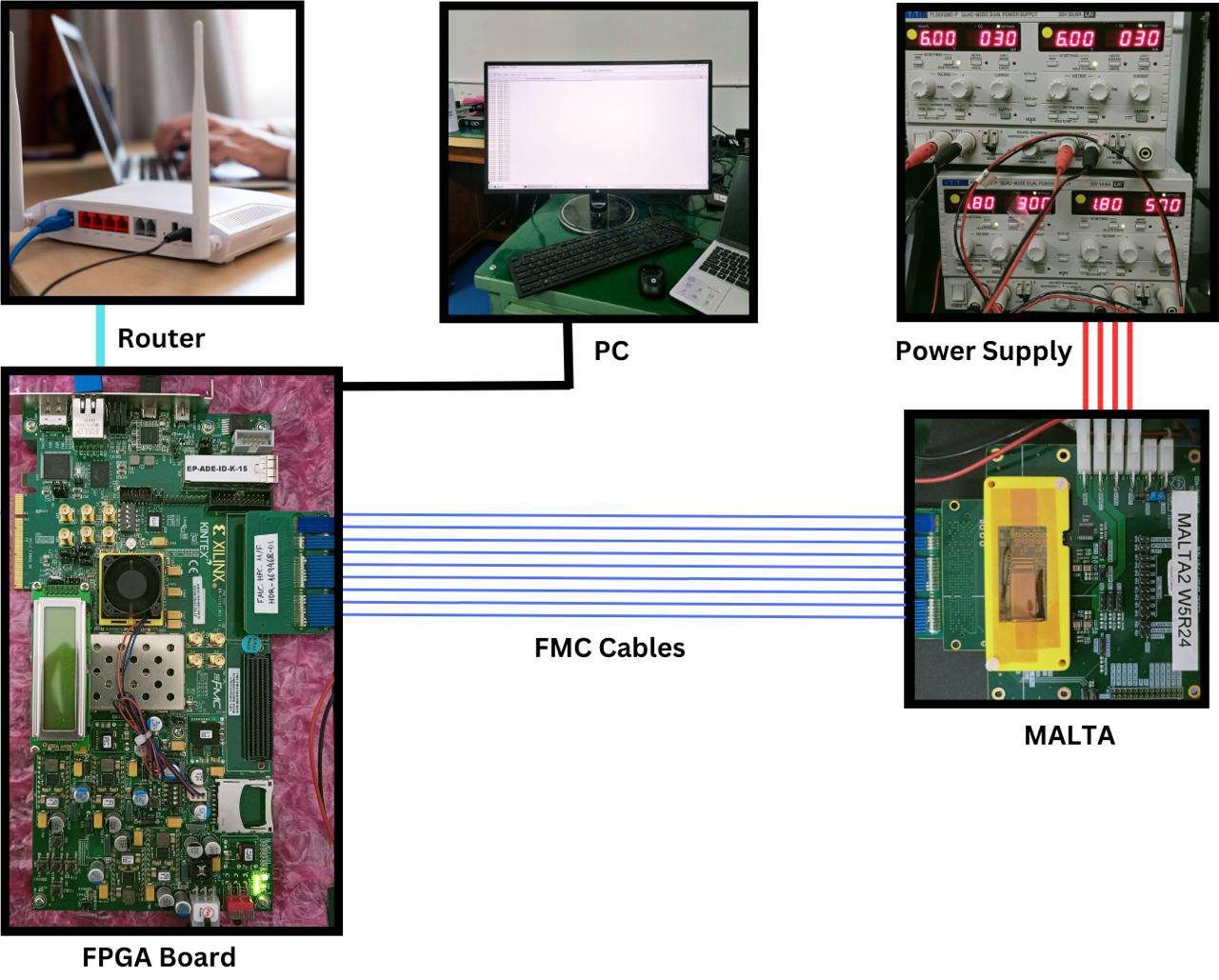}
    \label{sec1:setup}
}
\qquad
\subfigure[]{
    \includegraphics[width=.4\textwidth]{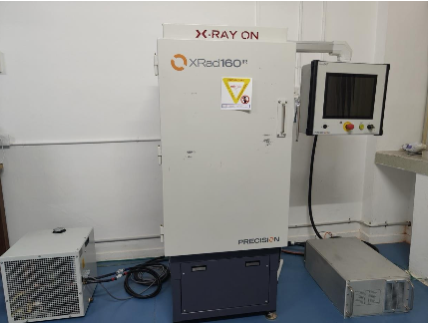}
    \label{sec2:xray}
}
\caption{(a) Characterization setup at IITM and (b) xRAD 160 x-ray irradiator \label{fig:1}}
\end{figure}

The MALTA2 chip is wire-bonded to a carrier board, which supplies it with various voltage levels. These include LVDD, DVDD, AVDD, PWELL, SUB, and DREF. The PWELL voltage is consistently maintained at -6V, while the reverse bias voltage (SUB) can be increased to generate a stronger drift field. However, the maximum bias voltage is limited by a compliance threshold of 2 mA, set as a safeguard to protect the electronics. LVDD, DVDD, AVDD and DREF are set at 1.8 V.  In total, four low-voltage power supplies are required to power up the MALTA2 sensor. \\
The carrier board interfaces with a Xilinx 7-series (KC705) FPGA board, which is responsible for reading out the chip data. This readout process utilizes a custom firmware project and communicates with the FPGA via Ethernet. The connection between the carrier board and the FPGA is established using an FMC cable. A computer is used to log and analyze the data from the sensor. Figure \ref{sec1:setup} shows the MALTA2 sensor characterization setup in IITM. For the irradiation study, the sensor is irradiated using X-ray to a dose of $10^{4}$ Gy. The irradiation is done using the xRAD 160 x-ray irradiator, which is shown in Figure \ref{sec2:xray}. The irradiator uses a tungsten target and is capable of generating x-rays up to 160 kVp. 

\section{Characterization of MALTA2 sensor}
\label{sec:characterization}
MALTA2 is widely studied for its electrical properties, radiation hardness, and timing performance. This study focuses on the analysis of key DAC parameters in two distinct MALTA2 chips: one featuring the NAGAP flavor with a 300 $\mu$m substrate thickness, and the other utilizing the XDPW flavor with a 100 $\mu$m substrate thickness. Both the chips are produced on epitaxial substrate and with H-dop $n^{-}$ layers. They both operate at a substrate and p-well voltage of -6V and are maintained at a controlled temperature of 16 $^{\circ}$C. The analog front-end architecture of each pixel in the MALTA2 sensor \cite{Piro} incorporates several DAC settings that enable fine-tuning of the front-end response for optimal sensor performance. These include IBIAS, IDB, ITHR, ICASN, VRESETD, VRESETP, VCASN, and VCLIP, each playing a distinct role in defining the signal processing characteristics.
In this study, the sensor response is investigated by systematically varying three critical DAC parameters: IDB, ITHR, and ICASN, while keeping all other DAC settings constant. The IDB DAC primarily determines the threshold of the discriminator circuit, directly influencing the comparator's response to the incoming signal. The ITHR DAC plays a crucial role in shaping the pulse duration and controlling the amplifier gain, which in turn affects signal clarity, noise rejection, and overall front-end performance. Meanwhile, the ICASN DAC regulates the baseline voltage of the amplifier, thereby defining the operating point of the front-end circuit. To systematically evaluate the impact of these DAC parameters on the pixel threshold, their respective current ranges are adjusted within their maximum limits: 200 $\mu$A for IDB, 210 $\mu$A for ITHR, and 20 $\mu$A for ICASN. 

\begin{figure}[htbp]
\centering
\subfigure[]{
    \includegraphics[width=.45\textwidth]{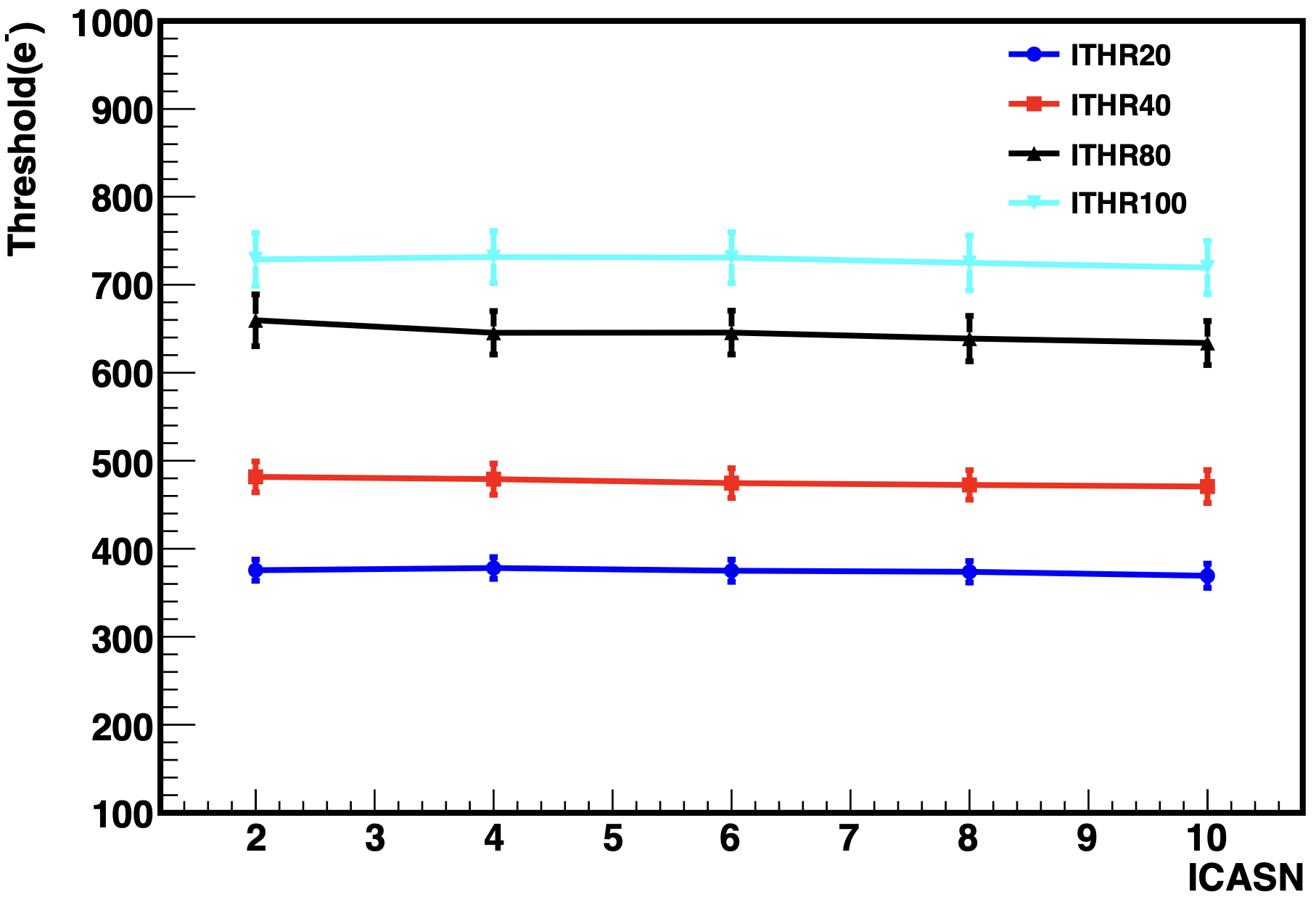}
    \label{fig:2a}
}
\qquad
\subfigure[]{
    \includegraphics[width=.45\textwidth]{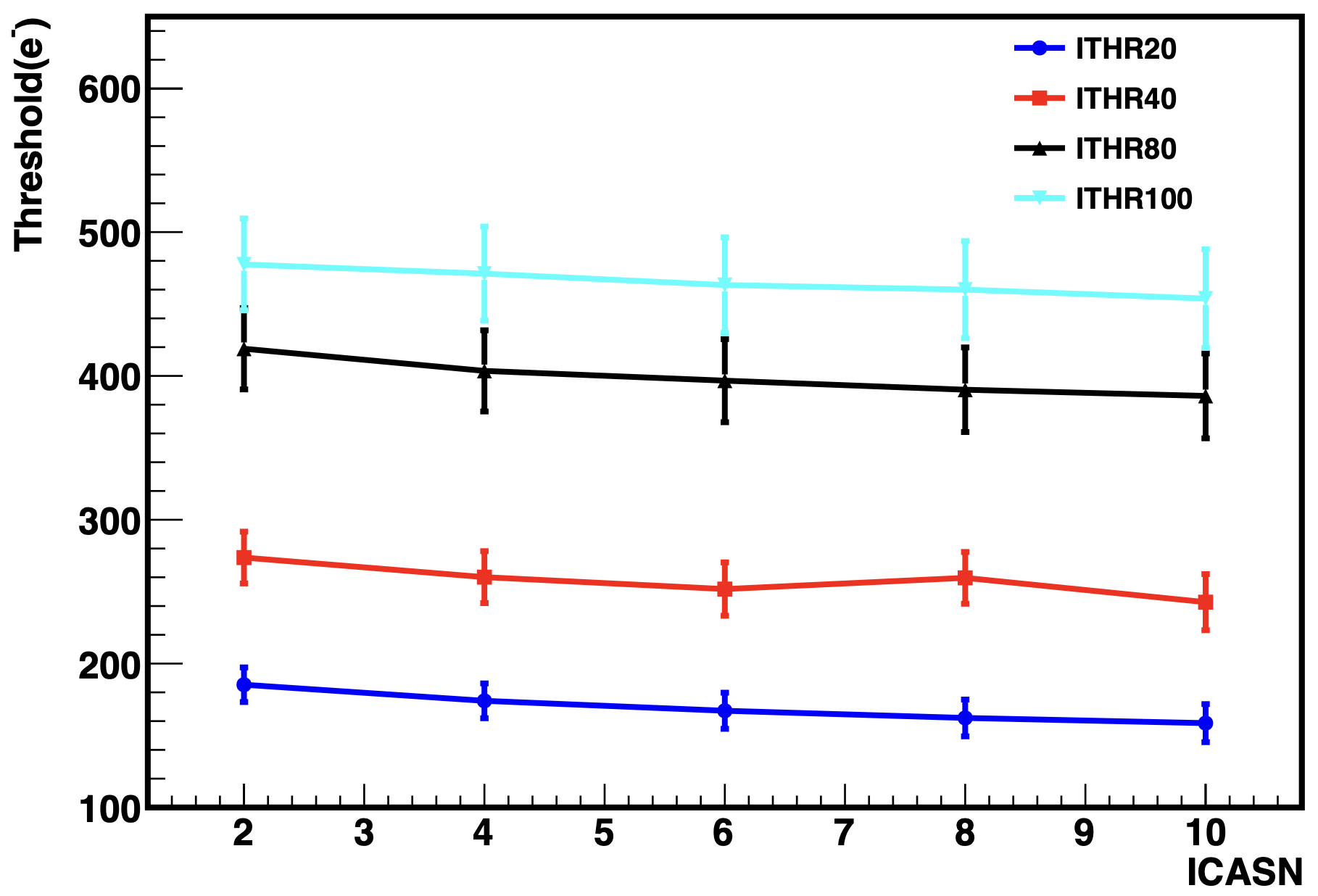}
    \label{fig:2b}
}
\caption{Threshold vs ICASN distribution for Various ITHR Values in (a) NGAP 300 $\mu$m chip and (b) XDPW 100 $\mu$m chip.\label{fig:2}}
\end{figure}
Figure \ref{fig:2a} illustrates the threshold distribution as a function of ICASN for different ITHR values in a non-irradiated 300 $\mu$m chip. The results indicate that the threshold increases with rising ITHR, while it remains relatively stable across varying ICASN values. Similarly, Figure \ref{fig:2b} presents the threshold distribution as a function of ICASN for a non-irradiated 100 $\mu$m chip. In this case, the threshold also increases with increasing ITHR; however, the threshold exhibits a slight decrease as ICASN increases. In both cases, the IDB parameter is maintained at a constant value of 100.

\section{Irradiation results}
\label{sec:results}
\begin{figure}[htbp]
\centering
\subfigure[]{
    \includegraphics[width=.45\textwidth]{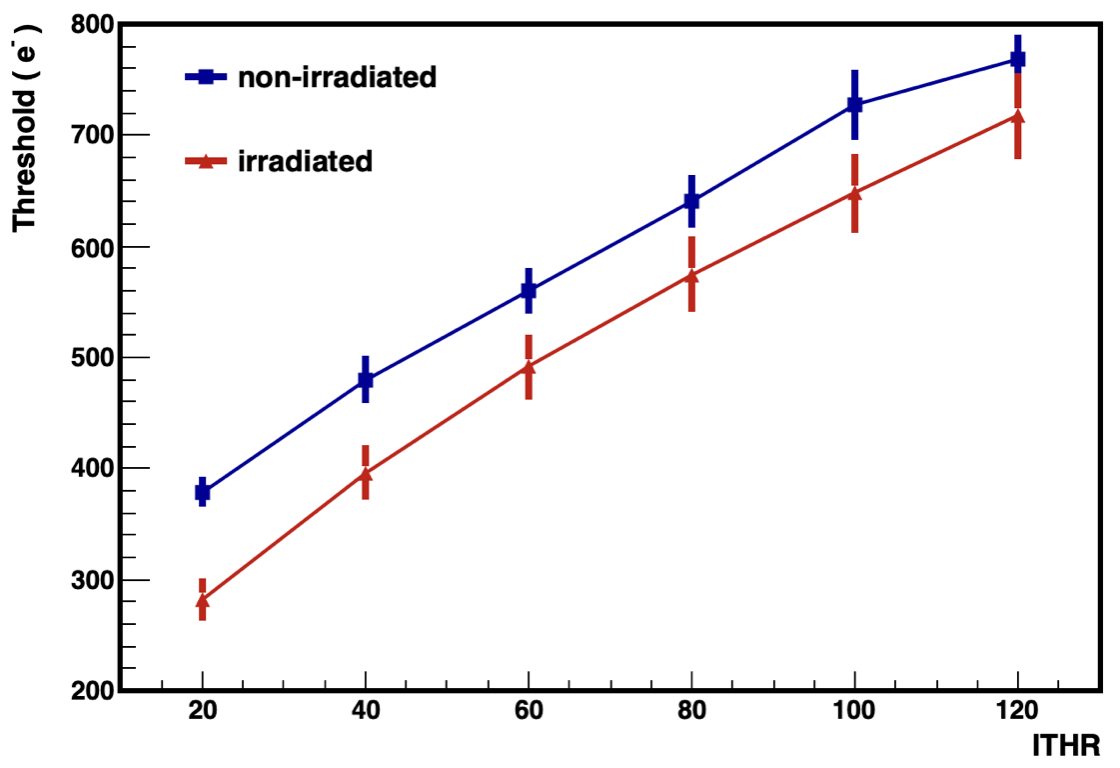}
    \label{fig:3a}
}
\qquad
\subfigure[]{
    \includegraphics[width=.45\textwidth]{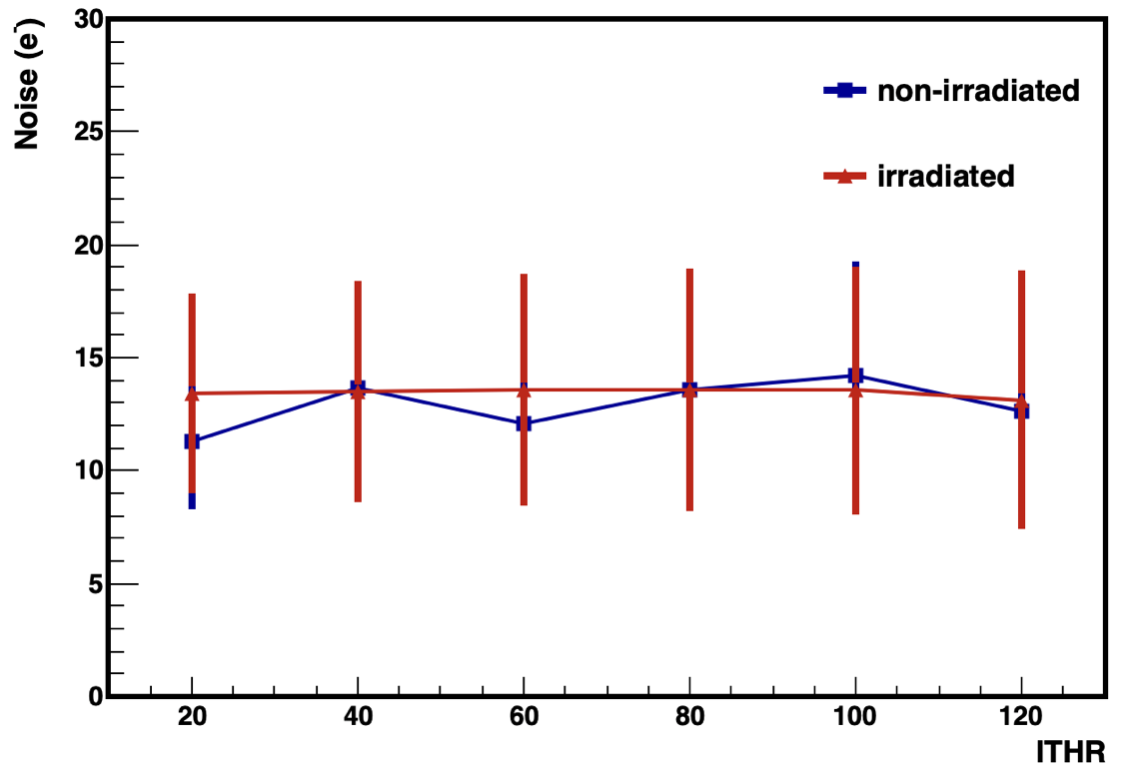}
    \label{fig:3b}
}
\caption{Comparison of (a) Threshold vs ITHR distribution and (b) Noise vs ITHR distribution between non-irradiated and irradiated MALTA2 Chips. \label{fig:3}}
\end{figure}

To investigate the effects of irradiation on the MALTA2 sensor, an NGAP-flavored sensor with a thickness of 300 $\mu m$ was exposed to X-ray irradiation at a total dose of $10^{4}$ Gy. The sensor's response was systematically evaluated both before and after irradiation to assess potential performance variations. Figure \ref{fig:3a} presents a comparative analysis of the threshold values of the sensor before and after irradiation for different values of ITHR. The results indicate a reduction in threshold following X-ray exposure. Figure \ref{fig:3b} illustrates the noise levels of the sensor under identical conditions. The results indicate that the noise level remains unchanged after irradiation, showing no adverse effects from X-ray exposure. Moreover, the noise remains stable across different ITHR settings. 

\section{Conclusion}
\label{sec:conclusion}
In this study, threshold measurements were performed on MALTA2 chips across various DAC settings, with results aligning well with design expectations. A comparative analysis of threshold and noise levels before and after X-ray irradiation confirmed that irradiation leads to a reduction in threshold, consistent with predictions. Meanwhile, the noise level remained stable, indicating that irradiation does not introduce additional noise. These findings demonstrate the robustness of the MALTA2 sensor in maintaining noise performance despite irradiation effects. Future work will focus on assessing sensor performance under higher X-ray irradiation doses to further understand its radiation tolerance and operational limits.

\acknowledgments
The project has been funded by the Science and Engineering Research Board (SERB), India, and the Ministry of Human Resource Development (MHRD), India.




\end{document}